\documentstyle[12pt,aps]{revtex}

\begin{document}
\title{Is the Space-Time Non-commutativity Simply Non-commutativity of Derivatives?\thanks{Talk given at the 8th International  Conference on Squeezed States and Uncertainty Relations (ICSSUR'2003), Puebla, Pue., M\'exico, June 9-13, 2003.}}
\author{Valeri V. Dvoeglazov}
\address{Universidad de Zacatecas\\
Apartado Postal 636, Suc. UAZ\\
Zacatecas 98062, Zac., M\'exico\\
E-mail: valeri@ahobon.reduaz.mx}


\maketitle

\begin{abstract}
Recently, some problems have been found in the definition of the partial derivative in the case of  the presence of both explicit and implicit
functional dependencies in the classical analysis. In this talk we investigate the influence of this observation on the quantum mechanics and classical/quantum  field theory. Surprisingly, some commutators of the coordinate-dependent operators  are not equal to zero. Therefore, we try to provide mathematical  foundations to the modern non-commutative theories. We also indicate
possible applications in the Dirac-like theories.
\end{abstract}

The assumption that operators of
coordinates do {\it not} commute $[\hat{x}_{\mu },\hat{x}_{\nu }]_{-} = i\theta_{\mu\nu}$ (or, alternatively, $[\hat{x}_{\mu },\hat{x}_{\nu }]_{-}= iC_{\mu\nu}^\beta x_\beta$)
has been first made by H. Snyder~\cite{snyder}. Later it was shown that such an anzatz may lead to non-locality. Thus, the Lorentz symmetry may be
broken. Recently, some attention has again been paid to this 
idea~\cite{noncom} in the context of ``brane theories''.

On the other hand, the famous Feynman-Dyson proof of Maxwell equations~\cite{FD}
contains intrinsically the non-commutativity of velocities. While $[ x^i, x^j ]_-=0$ therein, but  $[ \dot x^i (t), 
\dot x^j (t) ]_- = {i\hbar\over m^2} \epsilon^{ijk} B_k \neq 0$ (at the same time with $[x^i, \dot x^j]_- = {i\hbar \over m} \delta^{ij}$) that also may be considered as a contradiction with
the well-accepted theories. Dyson wrote in a very clever way: ``Feynman in 1948 was not alone in trying to build theories outside the framework of conventinal physics... All these radical programms, including Feynman's, failed... I venture to disagree with Feynman now, as I often did while he was alive..."

Furthermore, it was recently shown that notation and terminology, 
which physicists used when speaking about partial
derivative of many-variables functions, are sometimes 
confusing~\cite{chja} (see also the discussion in~\cite{eld}).
They referred to books~\cite{Arnold}: ``...one identifies sometime $f_1$ and $f$, saying, that is the same function represented with the help of variables $x_1$ instead of $x$. Such a simplification is very dangerous and may result in very serious contradictions" (see the text after Eq. (1.2.5) in~[6b]; $f=f(x)$, $f_1 = f (u(x_1))$). In~\cite{chja} the basic question was: how should one define correctly the time derivatives of the  functions $E [ x_1(t), \ldots x_{n-1} (t), t ]$ and $E ( x_1, \ldots x_{n-1}, t )$? Is there any sense in 
${\partial \over \partial t} E [{\bf r} (t), t]$ and ${d\over dt} 
E ({\bf r}, t)$?\footnote{The quotation from [4c, p. 384]: ``the [above] symbols
are meaningless, because the process denoted by the operator of {\it partial} differentiation can be applied only to functions of several {\it independent} variables and ${\partial \over \partial t} E [{\bf r} (t), t]$
is not {\it such} a function."} Those authors claimed that even well-known formulas 
\begin{equation}
{df \over dt} = \{ {\cal H}, f \} +{\partial f \over \partial t}\,,\quad
\mbox{and}\,\quad
{dE \over dt} = ({\bf v} \cdot {\bf \nabla} ) E + {\partial E\over \partial t}\,
\label{oldeq}
\end{equation}
can be confusing unless additional definitions present.\footnote{As for these formulas the authors of~\cite{chja} write:``this equation [cannot be correct] because the partial differentiation would involve increments of the functions ${\bf r} (t)$ in the form ${\bf r} (t) +\Delta {\bf r} (t)$ and we do not know how we must interpret this increment because we have two options: {\it either} $\Delta {\bf r} (t) = {\bf r} (t) - {\bf r}^\ast (t)$, {\it or} $\Delta {\bf r} (t) = {\bf r} (t) - {\bf r} (t^\ast)$. Both are different processes because the first one involves changes in the functional form of the functions ${\bf r} (t)$, while the second involves changes in the position along the path defined by ${\bf r} = {\bf r} (t)$ but preserving the same functional form." Finally, they gave the correct form, in their opinion, of (\ref{oldeq}). See in~[4d].}

Another well-known physical example of the situation, when we have both explicite and implicite dependences of the function which derivatives act upon, is the field of an accelerated charge~ \cite{landau}.
First, Landau and Lifshitz wrote that the functions depended on the retarded time $t^{\prime }$
and only through $t^{\prime }+R(t^{\prime })/c=t$ they depended implicitly
on $x,y,z,t$. However, later they used
the explicit dependence of $R$ and fields on the space coordinates of the
observation point too. Of course! Otherwise, the ``simply" retarded fields do not satisfy the Maxwell equations~[4b]. In the same work Chubykalo and Vlayev claimed that the time derivative
and curl did {\it not} commute in their case. Jackson, in fact, disagreed
with their claim on the basis of the definitions (``the equations
representing Faraday's law and the absence of magnetic charges ... are satisfied automatically''; see his Introduction in~[5b]). But,
he agrees with~\cite{landau} that one should find ``a contribution to the
spatial partial derivative for fixed time $t$ from explicit spatial
coordinate dependence (of the observation point)''. So, actually
the fields and potentials are the functions of the following forms:
$A^\mu (x, y, z, t' (x,y,z,t)), {\bf E} (x, y, z, t' (x,y,z,t)), {\bf B} (x, y, z, t' (x,y,z,t))$. The convincing abilities of Dr. Jackson are excellent! \v{S}kovrlj and Ivezi\'{c}~[5c] call this partial derivative as `{\it complete} partial
derivative'; Chubykalo and Vlayev~[4b], as `{\it total} derivative with respect to a given variable'; the terminology suggested by Brownstein~[5a] is
`the {\it whole}-partial derivative'. We shall denote below this whole-partial derivative operator as $\hat\partial \over \hat \partial 
x^i$, while still keeping the definitions of~[4c,d].

In~[5d] I studied the case when we deal with explicite and implicite dependencies  $f ({\bf p}, E ({\bf p}))$. It is well known that the energy in the
relativism is connected with the 3-momentum as $E=\pm \sqrt{{\bf p}^2 +m^2}$
; the unit system $c=\hbar=1$ is used. In other words, we must choose the
3-dimensional hyperboloid from the entire Minkowski space and the energy is 
{\it not} an independent quantity anymore. Let us calculate the commutator
of the whole derivative $\hat\partial /\hat\partial E$ and $\hat\partial / 
\hat\partial p_i$.\footnote{
In order to make distinction between differentiating the explicit function
and that which contains both explicit and implicit dependencies, the `whole
partial derivative' may be denoted as $\hat\partial$.} In the general case
one has 
\begin{equation}
{\frac{\hat\partial f ({\bf p}, E({\bf p})) }{\hat\partial p_i}} \equiv {
\frac{\partial f ({\bf p}, E({\bf p})) }{\partial p_i}} + {\frac{\partial f (
{\bf p}, E({\bf p})) }{\partial E}} {\frac{\partial E}{\partial p_i}}\, .
\end{equation}
Applying this rule, we surprisingly find 
\begin{eqnarray}
&&[{\frac{\hat\partial }{\hat\partial p_i}},{\frac{\hat\partial }{\hat
\partial E}}]_- f ({\bf p},E ({\bf p})) = {\frac{\hat\partial }{\hat\partial
p_i}} {\frac{\partial f }{\partial E}} -{\frac{\partial }{\partial E}} ({
\frac{\partial f}{\partial p_i}} +{\frac{\partial f}{\partial E}}{\frac{
\partial E}{\partial p_i}}) =  \nonumber \\
&=& {\frac{\partial^2 f }{\partial E\partial p_i}} + {\frac{\partial^2 f}{
\partial E^2}}{\frac{\partial E}{\partial p_i}} - {\frac{\partial^2 f }{
\partial p_i \partial E}} - {\frac{\partial^2 f}{\partial E^2}}{\frac{
\partial E}{\partial p_i}}- {\frac{\partial f }{\partial E}} {\frac{\partial
}{\partial E}}({\frac{\partial E}{\partial p_i}})\,.  \label{com}
\end{eqnarray}
So, if $E=\pm \sqrt{m^2+{\bf p}^2}$ 
and one uses the generally-accepted 
representation form of $\partial E/\partial p_i
=  p^i/E$,
one has that the expression (\ref{com})
appears to be equal to $(p_i/E^2) {\frac{\partial f({\bf p}, E ({\bf p}))}{
\partial E}}$. Within the choice of the normalization the coefficient is the
longitudinal electric field in the helicity basis (the electric/magnetic
fields can be derived from the 4-potentials which have been presented in~ 
\cite{hb}).\footnote{They are written in the following way:
\begin{eqnarray} &&\epsilon
_{\mu }({\bf p},\lambda =+1)={\frac{1}{\sqrt{2}}}{\frac{e^{i\phi }}{
p}}\pmatrix{ 0, {p_x p_z -ip_y p\over \sqrt{p_x^2 +p_y^2}}, {p_y p_z +ip_x
p\over \sqrt{p_x^2 +p_y^2}}, -\sqrt{p_x^2 +p_y^2}}\,, \\
&&\epsilon _{\mu }({\bf p},\lambda =-1)={\frac{1}{\sqrt{2}}}{\frac{e^{-i\phi }
}{p}}\pmatrix{ 0, {-p_x p_z -ip_y p\over \sqrt{p_x^2 +p_y^2}}, {-p_y p_z
+ip_x p\over \sqrt{p_x^2 +p_y^2}}, +\sqrt{p_x^2 +p_y^2}}\,, \\
&&\epsilon _{\mu }({\bf p},\lambda =0)={\frac{1}{m}}\pmatrix{ p, -{E \over p}
p_x, -{E \over p} p_y, -{E \over p} p_z }\,, \\
&&\epsilon _{\mu }({\bf p},\lambda =0_{t})={\frac{1}{m}}\pmatrix{E , -p_x,
-p_y, -p_z }\,.
\end{eqnarray}
And,
\begin{eqnarray}
&&{\bf E}({\bf p},\lambda =+1)=-{\frac{iEp_{z}}{\sqrt{2}pp_{l}}}{\bf p}-{\frac{
E}{\sqrt{2}p_{l}}}\tilde{{\bf p}},\quad {\bf B}({\bf p},\lambda =+1)=-{\frac{
p_{z}}{\sqrt{2}p_{l}}}{\bf p}+{\frac{ip}{\sqrt{2}p_{l}}}\tilde{{\bf p}}, \\
&&{\bf E}({\bf p},\lambda =-1)=+{\frac{iEp_{z}}{\sqrt{2}pp_{r}}}{\bf p}-{\frac{
E}{\sqrt{2}p_{r}}}\tilde{{\bf p}}^{\ast },\quad {\bf B}({\bf p},\lambda
=-1)=-{\frac{p_{z}}{\sqrt{2}p_{r}}}{\bf p}-{\frac{ip}{\sqrt{2}p_{r}}}\tilde{
{\bf p}}^{\ast }, \\
&&{\bf E}({\bf p},\lambda =0)={\frac{im}{p}}{\bf p},\quad {\bf B}({\bf p}
,\lambda =0)=0,
\end{eqnarray}
with $\tilde{{\bf p}}=\pmatrix{p_y\cr -p_x\cr -ip\cr}$. It is easy seen that
the parity properties of these vectors are different comparing with the standard basis. The parity operator for polarization vectors coincides with the metric tensor of
the Minkowski 4-space.} 
On the other hand, the commutator 
\begin{equation}
[{\frac{\hat\partial}{\hat\partial p_i}}, {\frac{\hat\partial}{\hat\partial
p_j}}]_- f ({\bf p},E ({\bf p})) = {\frac{1}{E^3}} {\frac{
\partial f({\bf p}, E ({\bf p}))}{\partial E}} [p_i, p_j]_-\,.
\end{equation}
This may be considered to be zero unless we would trust to the genious
Feynman. He postulated that the velocity (or, of course, the 3-momentum)
commutator is equal to $[p_i,p_j]\sim i\hbar\epsilon_{ijk} B^k$, i.e., to
the magnetic field.

Furthermore, since the energy derivative corresponds to the operator of time
and the $i$-component momentum derivative, to $\hat x_i$, we put forward the
following anzatz in the momentum representation: 
\begin{equation}
[\hat x^\mu, \hat x^\nu]_- = \omega ({\bf p}, E({\bf p})) \,
F^{\mu\nu}_{\vert\vert}{\frac{\partial }{\partial E}}\,,
\end{equation}
with some weight function $\omega$ being different for different choices of
the antisymmetric tensor spin basis. In the modern literature, the idea of the broken Lorentz invariance by this method is widely discussed, see e.g.~\cite{amelino}. 

Let us turn now to the application of the presented ideas to the Dirac case.
Recently, we analized Sakurai-van der Waerden method of derivations of the Dirac
(and higher-spins too) equation~\cite{Dvoh}. We can start from
\begin{equation}
(E I^{(2)}-{\bf \sigma}\cdot {\bf p}) (E I^{(2)}+ {\bf\sigma}\cdot
{\bf p} ) \Psi_{(2)} = m^2 \Psi_{(2)} \,,
\end{equation}
or
\begin{equation}
(E I^{(4)}+{\bf \alpha}\cdot {\bf p} +m\beta) (E I^{(4)}-{\bf\alpha}\cdot
{\bf p} -m\beta ) \Psi_{(4)} =0.\label{f4}
\end{equation}
Of course, as in the original Dirac work, we have
\begin{equation}
\beta^2 = 1\,,\quad
\alpha^i \beta +\beta \alpha^i =0\,,\quad
\alpha^i \alpha^j +\alpha^j \alpha^i =2\delta^{ij} \,.
\end{equation}
For instance, their explicite forms can be chosen 
\begin{eqnarray}
\alpha^i =\pmatrix{\sigma^i& 0\cr
0&-\sigma^i\cr}\,,\quad
\beta = \pmatrix{0&1_{2\times 2}\cr
1_{2\times 2} &0\cr}\,,
\end{eqnarray}
where $\sigma^i$ are the ordinary Pauli $2\times 2$ matrices.

We also postulate the non-commutativity
\begin{equation}
[E, {\bf p}^i]_- = \Theta^{0i} = \theta^i,,
\end{equation}
as usual. Therefore the equation (\ref{f4}) will {\it not} lead
to the well-known equation $E^2 -{\bf p}^2 = m^2$. Instead, we have
\begin{equation}
\left \{ E^2 - E ({\bf \alpha} \cdot {\bf p})
+({\bf \alpha} \cdot {\bf p}) E - {\bf p}^2 - m^2 - i {\bf\sigma}\times I_{(2)}
[{\bf p}\times {\bf p}] \right \} \Psi_{(4)} = 0
\end{equation}
For the sake of simplicity, we may assume the last term to be zero. Thus we come to
\begin{equation}
\left \{ E^2 - {\bf p}^2 - m^2 -  ({\bf \alpha}\cdot {\bf \theta})
\right \} \Psi_{(4)} = 0\,.
\end{equation} 
However, let us make the unitary transformation. It is known~\cite{Berg}
that one can\footnote{Of course, the certain relations for the components ${\bf a}$ should be assumed. Moreover, in our case ${\bf \theta}$ should not depend on $E$ and ${\bf p}$. Otherwise, we must take the noncommutativity $[E, {\bf p}^i]_-$ again.}
\begin{equation}
U_1 ({\bf \sigma}\cdot {\bf a}) U_1^{-1} = \sigma_3 \vert {\bf a} \vert\,.\label{s3}
\end{equation}
For ${\bf \alpha}$ matrices we re-write (\ref{s3}) to
\begin{eqnarray}
U_1 ({\bf \alpha}\cdot {\bf \theta}) U_1^{-1} = \vert {\bf \theta} \vert
\pmatrix{1&0&0&0\cr
0&-1&0&0\cr
0&0&-1&0\cr
0&0&0&1\cr} = \alpha_3 \vert {\bf\theta}\vert\,.
\end{eqnarray}
applying the second unitary transformation:
\begin{eqnarray}
U_2 \alpha_3 U_2^\dagger =
\pmatrix{1&0&0&0\cr
0&0&0&1\cr
0&0&1&0\cr
0&1&0&0\cr} \alpha_3 \pmatrix{1&0&0&0\cr
0&0&0&1\cr
0&0&1&0\cr
0&1&0&0\cr} = \pmatrix{1&0&0&0\cr
0&1&0&0\cr
0&0&-1&0\cr
0&0&0&-1\cr}\,.
\end{eqnarray}
The final equation is
\begin{equation}
[E^2 -{\bf p}^2 -m^2 -\gamma^5_{chiral} \vert {\bf \theta}\vert ] \Psi^\prime_{(4)} = 0\,.
\end{equation}
In the physical sense this implies the mass splitting for a Dirac particle over the non-commutative space. This procedure may be attractive for explanation of the mass creation and the mass splitting for fermions.

\bigskip
\bigskip

The presented ideas permit us to provide some
foundations for non-commutative field theories and induce us to look for further
applications of the functions with explicit and implicit dependencies in physics and mathematics.

\section*{Acknowledgments}
I am grateful to Profs. A. E. Chubykalo, R. Flores Alvarado, L. M. Gaggero Sager, V. Onoochin, M. Plyushchay and S. Vlayev  for discussions. 


\end{document}